\documentclass[twocolumn,preprintnumbers,amsmath,amssymb]{revtex4}
\usepackage{amsmath,amsfonts,latexsym,amssymb,graphicx,graphics,epsfig,subfigure,color,makeidx}
\usepackage{xcolor,diagbox}
\usepackage{multirow}
\usepackage[colorlinks,linkcolor=blue,anchorcolor=blue,citecolor=green,urlcolor=blue]{hyperref}
\usepackage{mathrsfs}
\usepackage{amsmath}
\newcommand {\nn}{\nonumber}

\begin{document}
\title{Graviscalar quasinormal modes and asymptotic tails of a thick brane}

\author{Qin Tan$^{a}$}
\author{Sheng Long$^{a}$}
\author{Weike Deng$^{a}$}
\author{Jiliang Jing$^{a}$}\email{jljing@hunnu.edu.cn, corresponding author.}

\affiliation{$^{a}$Department of Physics, Key Laboratory of Low Dimensional Quantum Structures and Quantum Control of Ministry of Education, Synergetic Innovation Center for Quantum Effects and Applications, Hunan Normal University, Changsha, 410081, Hunan, China}

\begin{abstract}
In this work, we investigate the quasinormal modes (QNMs) and the late time oscillatory tails of graviscalar perturbations in a thick brane. By considering the scalar perturbations of the thick brane metric, we obtain the main equations of graviscalar Kaluza-Klein modes. Using these equations, the frequencies of the graviscalar QNMs of the thick brane are obtained by the Wentzel-Kramers-Brillouin, asymptotic iteration, and numerical evolution methods. The results show that the scalar fluctuation of the thick brane has a series of discrete QNMs. In the four-dimensional spacetime, these modes manifest as decaying massive scalar particles. We also study in detail the late time tails of the scalar fluctuation. We find that the brane has slowly decaying oscillatory tails that may be new sources of the gravitational wave backgrounds. These novel phenomena may have potential phenomenological significance, and signals from these extra dimensions may potentially be found in current or future gravitational wave detectors.

\end{abstract}

\maketitle

\section{Introduction}
Since the 20th century, the idea that our universe might be embedded in a higher-dimensional spacetime as a four-dimensional object has sparked widespread interest in the spacetime view. The early motivation behind extra-dimensional theories was to unify various interactions in nature, such as the Kaluza-Klein (KK) theory and string theory~\cite{kaluza:1921un,Klein:1926tv,Scherk:1974ca}. With the development of extra-dimensional theories, some models focusing on addressing prominent issues in physics emerged, such as the Randall-Sundrum (RS) curved extra-dimensional model~\cite{Randall:1999ee,Randall:1999vf}. To explain the longstanding hierarchy problem in particle physics, namely the huge disparity between the weak scale and the Planck scale, Randall and Sundrum proposed the RS-I brane model composed of two branes embedded in a five-dimensional anti-de Sitter spacetime. Subsequently, based on the RS-I two-brane model, Randall and Sundrum further proposed the RS-II model with an infinite extra dimension. The inspiration behind the RS-II model was that even with an infinite extra dimension, the branes could still recover a four-dimensional effective gravitational potential. Due to this novel and intriguing characteristic, the RS-II model garnered widespread attention and has seen significant development in cosmology, particle physics, and black hole physics~\cite{Shiromizu:1999wj,Tanaka:2002rb,Zhidenko:2008fp,Gregory:2008rf,Konoplya:2011qq,Bauer:2016lbe,Jaman:2018ucm,Adhikari:2020xcg,Bhattacharya:2021jrn,Geng:2020fxl,Geng:2021iyq,Geng:2022dua}.

In the RS-II braneworld model, our world is embedded as a four-dimensional hypersurface in a five-dimensional spacetime, with the energy density along the extra dimension distributed as a delta function, hence it is termed the brane model. The thick brane model, which corresponds to this, is a smooth extension of the RS-II brane model, possessing a rich internal structure and thus has richer properties~\cite{DeWolfe:1999cp,Gremm:1999pj,Csaki:2000fc}. In thick brane theories, the localization of the gravitational zero mode and various matter fields is necessary. Various thick brane solutions of gravity theories and the localization of gravitational zero modes and matter fields on the brane have been studied~\cite{Afonso:2007gc,Dzhunushaliev:2010fqo,Dzhunushaliev:2011mm,Geng:2015kvs,Melfo2006,Almeida2009,Zhao2010,Chumbes2011,Liu2011,Bazeia:2013uva,Xie2017,Gu2017,ZhongYuan2017,ZhongYuan2017b,Zhou2018,Hendi:2020qkk,Xie:2021ayr,Moreira:2021uod,Xu:2022ori,Silva:2022pfd,Xu:2022gth}. However, it is important to note that the localization of the zero modes of gravity and various matter fields is aimed at recovering effective physics on the four-dimensional brane, but this does not bring us new insights. To explore the effects of the extra dimension, we are more interested in the massive KK modes beyond these zero modes, namely the massive KK modes that contain information about the extra dimension.

It is generally believed that apart from the zero modes, thick branes also have a continuum of massive KK modes. Our recent studies have indicated that within these continuous KK modes, there exists a series of discrete gravitational quasinormal modes (QNMs)~\cite{Tan:2022vfe,Tan:2023cra,Jia:2024pdk,Tan:2024url}. The spectrum of these QNMs is entirely determined by the structure of the thick brane, and conversely, these modes also contain crucial information about the extra dimension. Previous studies of the QNMs of the braneworld only considered the transverse-traceless tensor fluctuations of the metric, corresponding to spin-2 gravitons on the brane. For the RS-II thin brane model, this is sufficient because, without considering material sources on the brane, the metric fluctuation of the RS-II thin brane consists only of the transverse-traceless tensor mode~\cite{Randall:1999vf}. However, for the thick brane scenario, the situation is quite different. Since thick branes are generated by a background scalar field coupled with gravity, their complete metric fluctuations of thick branes should include tensor fluctuations, vector fluctuations (graviphoton), and scalar fluctuations (graviscalar)~\cite{Giovannini:2001fh,Giovannini:2001xg,Giovannini:2001vt,Kobayashi:2001jd}. Vector fluctuations and scalar fluctuations contain information about the matter field that constitutes the thick brane, which may be the key to distinguishing between the thick brane model and the thin brane model. Additionally, scalar fluctuations have also been widely studied in other higher-dimensional spacetimes, such as scalar fluctuations of brane cosmology~\cite{Langlois:2000iu,Koyama:2004ap,Hiramatsu:2004aa,Brax:2004xh,Maartens:2010ar,Maier:2013gua,Jaman:2018ucm,Banerjee:2020uil,Ravanpak:2022awx} and graviscalar QNMs of higher-dimensional black holes~\cite{Cardoso:2003vt,Natario:2004jd,Konoplya:2008ix}.

For vector fluctuations, generally only non-localized zero mode exist, with no other modes. Tensor fluctuations exhibit a bound zero mode and an infinite number of massive KK modes. As for the scalar fluctuations of thick branes, in most cases, zero modes exist but cannot be bound to the brane, and the massive mode is similar to the case of tensor fluctuations. So, the question now is, can we know more about the scalar fluctuations of thick branes? Do scalar fluctuations also exhibit QNMs like tensor fluctuations? By employing the methods developed from studying the QNMs of thick brane tensor fluctuations, we will investigate the QNMs of scalar fluctuations of flat thick branes in this paper, aiming to understand more about scalar fluctuations.

This paper is organized as follows. In Sec.~\ref{sec:background}, we review the thick brane solution and its linear scalar perturbation. In Sec.~\ref{sec:qnms}, we investigate the graviscalar QNMs of the thick brane using the Wentzel-Kramers-Brillouin (WKB) approximation, the asymptotic iteration method in the frequency domain, and obtain the waveform of these modes by numerical evolution. Finally, the conclusions and the discussions are given in Sec.~\ref{sec:conclusions}.

\section{Thick brane solution and its scalar perturbation}
\label{sec:background}
In this section, we will review the thick brane solution and corresponding scalar perturbation of the metric. Starting from the Einstein-Hilbert action minimally coupled to a canonical scalar field, the action is given by
\begin{eqnarray}
	S=\int d^5x\sqrt{-g}\left(\frac{1}{2\kappa^{2}_{5}}R-\frac{1}{2}g^{MN}\partial_{M}
	\phi\partial_{N}\phi -V(\phi)\right),\label{action}
\end{eqnarray}
where $\kappa_{5}^{2}\equiv8\pi G_{5}$ is set to $\kappa_{5} = 1$ for convenience. The metric is assumed to be a static flat case
\begin{equation}
	ds^2=e^{2A(y)}\eta_{\mu\nu}dx^\mu dx^\nu+dy^2,
	\label{metric1}
\end{equation}
where $A(y)$ is the warp factor. By varying the above action~(\ref{action}) with respect to the metric and the scalar field $\phi$, the field equations are given by
\begin{eqnarray}
	R_{MN}-\frac{1}{2}Rg_{MN}&=&-\frac{1}{2}g_{MN}\left(\partial^{A}\phi\partial_{A}\phi-V(\phi)\right) \nonumber\\
	&&+\partial_{M}\phi\partial_{N}\phi,\label{field equation}\\
	g^{MN}\nabla_{M}\nabla_{N}\phi&=&\frac{\partial V(\phi)}{\partial\phi}.\label{motion equation}
\end{eqnarray}
Substituting the metric~\eqref{metric1} into the above field equations, we can obtain the specific dynamic equations 
\begin{eqnarray}
	6(\partial_{y}A)^2 +3\partial_{y}^{2}A&=&-\frac{1}{2}(\partial_{y}\phi)^{2}-V,  \label{EoMs1}\\
	6(\partial_{y}A)^2&=&\frac{1}{2}(\partial_{y}\phi)^2-V,  \label{EoMs2}\\
	\partial_{y}^{2}\phi+4(\partial_{y}A)\partial_{y}\phi&=&\frac{\partial V}{\partial\phi}.  \label{EoMsphi}
\end{eqnarray}
By using the superpotential method, the thick brane solution is given in Ref.~\cite{Gremm:1999pj}:
\begin{eqnarray}
	A(y)&=&-\ln\left(\cosh(ky)\right),\label{warpfactorsolution1}\\
	\phi(y)&=&\sqrt{3}\arcsin\left(\tanh\left(ky\right)\right),\label{scalarfieldsolution1}\\
	V(\phi)&=&\frac{3k^{2}}{4}\left(5\cos\left(\sqrt{\frac{4}{3}}\phi\right)-3\right),\label{scalarpotentialsolution1}
\end{eqnarray}
where $k$ is a constant with mass dimension one. 

To investigate the QNMs of scalar fluctuation for the thick brane, we next consider scalar perturbation of the metric. In braneworld models, to study the fluctuations of the metric and the field, it is generally necessary to transform the extra-dimensional coordinate $y$ to a conformal flat coordinate $z$, that is, to perform the following coordinate transformation:
\begin{eqnarray}
	dz=e^{-A(y)}dy,
\end{eqnarray}
then, metric~\eqref{metric1} becomes
\begin{equation}
	ds^2=e^{2A(z)}(\eta_{\mu\nu}dx^\mu dx^\nu+dz^2).
	\label{metric2}
\end{equation}
Since the linear perturbation of the metric of a braneworld can be decomposed into scalar, transverse vector, and transverse-traceless tensor modes, and these three modes are decoupled from each other after performing a scalar-tensor-vector decomposition. Therefore, we can consider the scalar perturbation of the metric separately. The perturbed metric in the longitudinal gauge is~\cite{Kobayashi:2001jd}
\begin{equation}
ds^2=e^{2A(z)}(1+2\varphi(x^{\mu},z))(\eta_{\mu\nu}dx^\mu dx^\nu+(1+2\Psi(x^{\mu},z))dz^2),
	\label{perturbedmetric}
\end{equation}
On the other hand, the perturbed background scalar field is $\phi_{0}=\phi(z)+\delta\phi(x^{\mu}+z)$. By substituting the perturbed metric and scalar field into Eqs.~\eqref{field equation} and ~\eqref{motion equation}, we obtain the equations of the scalar perturbation:

\begin{eqnarray}
(z,z):&&3\eta^{\alpha\beta}\partial_{\alpha}\partial_{\beta}\Psi+12\partial_{z}A\partial_{z}\Psi-12(\partial_{z}A)^2\partial_{z}\varphi \nonumber\\ 
&&=\partial_{z}\phi\partial_{z}\delta\phi-\varphi(\partial_{z}\delta\phi)^2-e^{2A}\frac{\partial V}{\partial \phi}\delta\phi,\label{zzcompont}\\
(z,\mu):&&-3\partial_{\mu}\partial_{z}\Psi+3\partial_{z}A\partial_{\mu}\varphi=\partial_{z}\phi\partial_{\mu}\delta\phi,\label{zmucompont}\\
(\mu,\nu):&&\Big(3\partial_{z}^{2}\Psi-6\partial_{z}^{2}A\varphi-3\partial_{z}A\partial_{z}\varphi+9\partial_{z}A\partial_{z}\Psi\nonumber\\&&-6(\partial_{z}A)^{2}\varphi+\eta^{\alpha\beta}\partial_{\alpha}\partial_{\beta}\varphi+2\eta^{\alpha\beta}\partial_{\alpha}\partial_{\beta}\Psi\Big)\delta_{\nu}^{\mu}\nonumber\\
&&-\eta^{\mu\beta}\partial_{\beta}\partial_{\nu}\varphi-2\eta^{\mu\beta}\partial_{\beta}\partial_{\nu}\Psi\nonumber\\
&&=\Big(-\partial_{z}\phi\partial_{z}\delta\phi+\varphi(\partial_{z}\delta\phi)^2-e^{2A}\frac{\partial V}{\partial \phi}\delta\phi\Big)\delta_{\nu}^{\mu},\nonumber\\ \label{munucompont}\\
\text{matter}:&&\partial_{z}^{2}\delta\phi+3\partial_{z}A\partial_{z}\delta\phi+(4\partial_{z}\Psi-\partial_{z}\varphi-6\partial_{z}\varphi)\partial_{z}\phi\nonumber \\ &&-2\varphi\partial_{z}^{2}\phi_{0}+\eta^{\alpha\beta}\partial_{\alpha}\partial_{\beta}\delta\phi=e^{2A}\frac{\partial V}{\partial \phi}\delta\phi.\label{mattercompont}
\end{eqnarray}
From the off-diagonal part of Eq.~\eqref{munucompont}, we can obtain
\begin{equation}
\varphi+2\Psi=0.\label{offdiagonalmunucompont}
\end{equation}
From Eq.~\eqref{zmucompont} we get
\begin{equation}
	\delta\phi=\frac{(-3\partial_{z}\Psi+3\partial_{z}A\varphi)}{\partial_{z}\phi}.\label{zmucompont1}
\end{equation}
This means that there is only one physical scalar degree of freedom. Combining the above equations~\eqref{zzcompont}-\eqref{offdiagonalmunucompont}, we find the master equation for the scalar perturbation is
\begin{eqnarray}
 \Box^{(4)}\Psi&&+\left(4\partial_{z}^{2}A-\frac{4\partial_{z}A\partial_{z}^{2}\phi}{\partial_{z}\phi}\right)\Psi\nonumber\\
 &&+\left(3\partial_{z}A-
 \frac{2\partial_{z}^{2}\phi}{\partial_{z}\phi}\right)\partial_{z}\Psi+\partial_{z}^{2}\Psi=0, \label{mainescalarquation}
\end{eqnarray}
where $\Box^{(4)}=\eta^{\alpha\beta}\partial_{\alpha}\partial_{\beta}$. Introducing the following Kaluza-Klein decomposition
\begin{equation}
	\Psi(x^{\mu},z)=e^{-\frac{3}{2}A(z)}\partial_{z}\phi \tilde{\Psi}(t,z) e^{-i a_{i}x^{i}},\label{decomposition1}
\end{equation}
we can obtain the wave equation:
\begin{equation}
	-\partial_{t}^{2}\tilde{\Psi}+\partial_{z}^{2}\tilde{\Psi}-U(z)\tilde{\Psi}-a^{2}\tilde{\Psi}=0, \label{evolutionequation}
\end{equation}
where
\begin{eqnarray}
	U(z)&=&-\frac{5}{2}\partial_{z}^{2}A+\frac{9}{4}(\partial_{z}A)^{2}-\frac{\partial_{z}^{3}\phi}{\partial_{z}\phi}\nn\\
	&&+\partial_{z}A\frac{\partial_{z}^{2}\phi}{\partial_{z}\phi}+2\left(\frac{\partial_{z}^{2}\phi}{\partial_{z}\phi}\right)^{2}\label{effectivepotential}
\end{eqnarray} 
is the effective potential. Parameter $a=\sqrt{a^{i}a_{i}}$ corresponds to the magnitude of the spatial three-momentum of the KK particle along the brane. When $a=0$, it means that the KK particle only moves along the extra dimension and moves at the speed of light at infinity in the extra dimension. When $a>0$, the graviscalar KK mode moves both along the extra dimension and on the brane. Then we decompose the function $\tilde{\Psi}$ 
further as $\tilde{\Psi}=e^{-i\omega t}\psi(z)$ and substitute it into the above wave equation~\eqref{evolutionequation}, a Schr\"odinger-like equation can be obtained as
\begin{equation}
	-\partial_{z}^{2}\psi(z)+U(z)\psi(z)=m^{2}\psi(z),\label{Schrodingerlikeequation}
\end{equation}
where $m=\sqrt{\omega^{2}-a^{2}}$ is the mass of the graviscalar KK mode. Unlike the case of the transverse-traceless tensor perturbation, we do not want the above equation to have a bound zero mode of scalar perturbation. Because if there is a scalar zero mode, there will be a ``fifth force'' on the brane, which is unacceptable. In order to obtain the solution of the scalar zero mode, we decompose the above equation into the following supersymmetric form
\begin{equation}
	H^{\dagger}H\phi(z)=m^{2}\psi(z)\label{supersymmetricform},
\end{equation}
where $H^{\dagger}$ and $H$ are
\begin{eqnarray}
	H^{\dagger}&=&-\partial_{z}+\frac{3}{2}\partial_{z}A(z)-\frac{\partial^{2}_{z}A(z)}{\partial_{z}A(z)}+\frac{\partial^{2}_{z}\phi(z)}{\partial_{z}\phi(z)},\\
	H&=&\partial_{z}+\frac{3}{2}\partial_{z}A(z)-\frac{\partial^{2}_{z}A(z)}{\partial_{z}A(z)}+\frac{\partial^{2}_{z}\phi(z)}{\partial_{z}\phi(z)}.
\end{eqnarray}
Thus, the scalar zero mode is $\psi_{0}\propto\frac{\partial_{z}A(z)}{e^{\frac{3}{2}A(z)}\partial_{z}\phi(z)}$. Generally, this scalar zero mode cannot be localized. In addition to the zero mode, the above equations also support massive KK modes. Due to the complexity of the effective potential, it is difficult to obtain an analytical solution of the massive KK mode. But from the perspective of the QNMs, we can gain a better understanding of these massive scalar KK modes.

\section{scalar quasinormal modes of brane}
\label{sec:qnms}
\begin{figure}[htbp]
 	\centering
 	\includegraphics[width=0.35\textwidth]{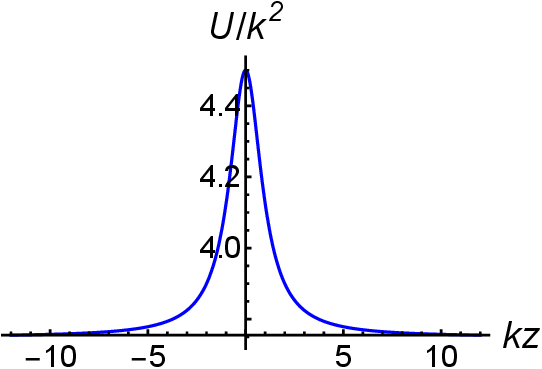}
 	\caption{The shape of the effective potential~(\ref{effectivepotentialz}).}\label{effectiveplot}
 \end{figure}

Now, we investigate the graviscalar QNMs of the thick brane. The warp factor~\eqref{warpfactorsolution1} and the background scalar field~\eqref{scalarfieldsolution1} are rewritten in terms of the $z$ coordinate as
\begin{eqnarray}
	A(z)&=&-\frac{1}{2}\ln(k^{2}z^{2}+1),\label{warpfactorsolutionz}\\
	\phi(z)&=&\sqrt{3}\arctan(kz).\label{scalarfieldsolutionz}
\end{eqnarray}
Substituting the above warp factor $A(z)$ and the background scalar field $\phi(z)$ into the effective potential ~\eqref{effectivepotential}, the specific form of $U(z)$ is given by
\begin{eqnarray}
	U(z)&=&\frac{3k^2(5k^2z^{2}+6)}{4(k^2z^{2}+1)^{2}}.\label{effectivepotentialz}
\end{eqnarray}
We plot the effective potential $U(z)$ in Fig.~\ref{effectiveplot}. We can see that the effective potential is a pure barrier and $U(\pm\infty)\rightarrow0$. In addition, the specific form of the scalar zero mode is
\begin{equation}
\psi_{0}(z)=\frac{\sqrt{3}z(1+z^{2})^{3/4}}{3}.\label{scalarzeromodez}
\end{equation}
Obviously, this zero mode cannot be localized on the brane. Now we turn our attention to the massive KK modes. Since the effective potential~\eqref{effectivepotentialz} approaches to $0$ at $kz=\pm\infty$, these scalar massive KK modes always tunnel to extra dimension infinity. Thus, the boundary conditions of these massive KK modes are
\begin{equation}
	\label{boundaryconditions1}
	\psi(z) \propto \left\{
	\begin{aligned}
		e^{im z}, &~~~~~z\to\infty.& \\
		e^{-im z},  &~~~~~z\to-\infty,&
	\end{aligned}
	\right.
\end{equation}
which is only an outgoing wave at the spatial infinity and only an ingoing wave at negative spatial infinity. With the equation and the boundary conditions, we can solve the graviscalar QNMs of the thick brane.
Since the effective potential is similar to the case of the gravitational perturbation of a black hole. Based on this similarity, we use some methods to solve the QNMs of black holes to solve the graviscalar QNMs of the thick brane. 

$WKB~method$ The WKB method is a very common method for solving the QNMs of a black hole~\cite{Iyer:1986np,Kanti:2006ua}. This method is based on matching the asymptotic WKB solutions at the event horizon and space infinity with the Taylor expansion near the peak of effective potential barrier through the two turning points. The complex frequency formula of the sixth order WKB method is~\cite{Konoplya:2003ii}
\begin{eqnarray}
i\frac{m^{2}-U_{0}}{\sqrt{-2U''_{0}}}-\sum_{j=2}^{6}\Lambda_{j}=n+\frac{1}{2},~~~n=1,~2~,3\dots,\label{WKBm}
\end{eqnarray}
where $U_{0}$ is the peak value of the effective potential $U(z)$, $U''_{0}$ is the second derivative with respect to the extra dimensional coordinate $z$ at the maximum of the potential, and $\Lambda_{j}$ are 2nd to 6th correction terms. The explicit form of $\Lambda_{j}$ can be found in Ref.~\cite{Konoplya:2003ii}. Using the WKB method, we obtain the first four scalar quasinormal frequencies (QNFs) of the brane, which can be seen from Tab.~\ref{tab1}. For higher overtones, the QNFs can not be solved by the WKB method. We need to use other methods.

\begin{table}[htbp]
	\begin{tabular}{|c|c|c|}
		\hline
		$\;\;n\;\;$  &
		$\;\;\text{WKB method}\;\;$  &
		$\;\;\;\;\;\;\;\;\text{AIM}\;\;\;\;\;\;\;$ \\
		\hline
		~   &~~~~$\text{Re}(m/k)$  ~~  $\text{Im}(m/k)~~$  &$~~~~~~\text{Re}(m/k)$ ~~ $\text{Im}(m/k)~~$       \\
		0   &2.000357~~ -0.508985       &~~1.999287~~ -0.507338       \\
		1   &1.754400~~ -1.571916      &~~1.747004~~ -1.576767  \\
		2   &1.263268~~ -2.833437        &~~1.276446~~ -2.862700   \\
		3   &0.610778~~ -4.548061        &~~0.839490~~ -4.476666   \\
		\hline
	\end{tabular}
	\caption{The first four QNFs using the WKB method and the AIM.\label{tab1}}
\end{table}

$Asymptotic~iteration~method$
In order to accurately solve high overtone modes, we use the asymptotic iteration method (AIM). The AIM is an effective method to find the eigenvalues of linear second-order homogeneous differential equation~\cite{Ciftci:2003As,ciftci:2005co}. In general, if the equation has the following form
\begin{equation}
	y''(x)=\lambda_{0}(x)y'(x)+s_{0}(x)y(x),\label{2orderdiffeq}
\end{equation}
where $\lambda_{0}(x)\neq0$ and $s_{0}(x)$ are smooth functions. Taking the differentiation of the above Eq.~\eqref{2orderdiffeq}, one  finds
\begin{equation}
	y'''(x)=\lambda_{1}(x)y'(x)+s_{1}(x)y(x),
\end{equation}
where
\begin{eqnarray}
	\lambda_{1}(x)&=&\lambda'_{0}+s_{0}+\lambda_{0}^{2},\\
	s_{1}(x)&=&s'_{0}+s_{0}\lambda_{0}.
\end{eqnarray}
Then, taking $n-$time derivatives of Eq.~\eqref{2orderdiffeq} we have
\begin{eqnarray}
	y^{(n+2)}(x)&=&\lambda_{n}(x)y'(x)+s_{n}(x)y(x),	
\end{eqnarray}
where
\begin{eqnarray}
	\lambda_{n}(x)&=&\lambda'_{n-1}+s_{n-1}+\lambda_{0}\lambda_{n-1}, \label{AIMrelation1}\\ s_{n}(x)&=&s'_{n-1}+s_{0}\lambda_{n-1}\label{AIMrelation2}.
\end{eqnarray}
With sufficiently large $n$, the asymptotic aspect of the AIM claims that
\begin{eqnarray}
	\frac{s_{n}(x)}{\lambda_{n}(x)}=\frac{s_{n-1}(x)}{\lambda_{n-1}(x)}=\beta(x),\label{QNMscondition1}
\end{eqnarray}
or an equivalent form
\begin{eqnarray}
	s_{n}(x)\lambda_{n-1}(x)-s_{n-1}(x)\lambda_{n}(x)=0.\label{QNMscondition2}
\end{eqnarray}
The QNFs can be solved from the above ``quantization condition". However, computing the recurrence relations~\eqref{AIMrelation1} and \eqref{AIMrelation2} requires a lot of resources. Thus, Cho $et~al$ improved the original AIM by using the Taylor series~\cite{Cho:2011sf}. This greatly improved the accuracy and speed of numerical calculation. Expanding the functions $\lambda_{n}(x)$ and $s_{n}(x)$ at the point $\chi$:
\begin{eqnarray}
	\lambda_{n}(x)&=&\sum_{i=0}^{\infty}c_{n}^{i}(x-\chi)^{i},\\
	s_{n}(x)&=&\sum_{i=0}^{\infty}d_{n}^{i}(x-\chi)^{i}.
\end{eqnarray}
where the $i$-th Taylor coefficients of $\lambda_{n}$ and $s_{n}$ are denoted by $c_{n}^{i}$ and $d_{n}^{i}$, respectively. Now, Eqs.~\eqref{AIMrelation1} and ~\eqref{AIMrelation2} become
\begin{eqnarray}
	c_{n}^{i}&=&(i+1)c_{n-1}^{i+1}+d^{i}_{n-1}+\sum_{k=0}^{i}c_{0}^{k}c_{n-1}^{i-k},\\
	d_{n}^{i}&=&(i+1)d_{n-1}^{i+1}+\sum_{k=0}^{i}d_{0}^{k}c_{n-1}^{i-k}.
\end{eqnarray}
The ``quantization condition''~\eqref{QNMscondition2} can be expressed in terms of these coefficients 
\begin{equation}
	d_{n}^{0}c_{n-1}^{0}-d_{n-1}^{0}c_{n}^{0}=0\label{QNMscondition3}.
\end{equation}
We can see that the final recurrence relations do not require derivative operations in the iteration process. We can obtain the QNFs by solving this simple recurrence relation. Obviously, the improved AIM depends on the expansion point $\chi$. It is found that the value of the expansion point has great influence on the convergence speed. Furthermore, the AIM seems to converge most quickly when $\chi$ is chosen to be the maximum of the effective potential. 

\begin{figure}[htbp]
	\centering
	\includegraphics[width=0.35\textwidth]{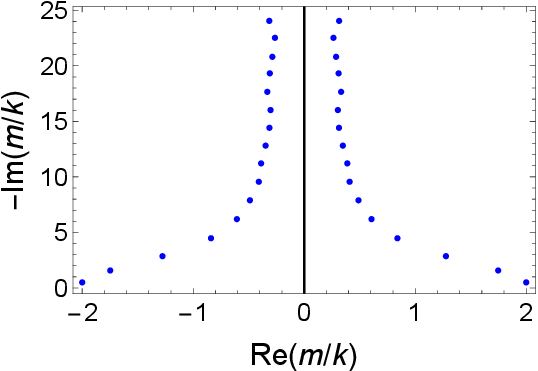}
	\caption{The first fifteen graviscalar QNFs of the thick brane solved by the improved AIM.}\label{figqnmsplot}
\end{figure}

Now, we use the AIM to solve the graviscalar QNMs of the thick brane. Since Eq.~\eqref{Schrodingerlikeequation} does not contain a first-order derivative term, we perform the coordinate  transformation $u= \frac{\sqrt{4k^2 z^2+1}-1}{2k z}$ to get a form that applies to the AIM. The Schr\"odinger-like~~\eqref{Schrodingerlikeequation} then becomes
\begin{eqnarray}
	\frac{\left(u^2-1\right)^3 \left(\left(u^4-1\right) \psi ''(u)+2 u \left(u^2+3\right) \psi
		'(u)\right)}{\left(u^2+1\right)^3}\nonumber\\
	+\left(\frac{m^2}{k^2}-\frac{3 \left(u^2-1\right)^2 \left(6 u^4-7 u^2+6\right)}{4
		\left(u^4-u^2+1\right)^2}\right) {\psi} (u)=0,\label{Schrodingerlikeequation1}
\end{eqnarray}
In this coordinate, the boundary conditions~\eqref{boundaryconditions1} are rewritten as
\begin{equation}
	\label{transformboundaryconditions}
\psi(u) \propto \left\{
	\begin{aligned}
		e^{-\frac{i m/k }{2 u-2}}, &~~~ u\to 1,& \\
		e^{\frac{i m/k }{2 u+2}}, &~~~ u\to -1.&
	\end{aligned}
	\right.
\end{equation}
To accommodate the above boundary conditions, we define
\begin{eqnarray}
	\psi(u)=\tilde{\psi} (u) e^{-\frac{i m/k }{2 u-2}} e^{\frac{i m/k }{2 u+2}}.\label{boundarysolutions}
\end{eqnarray}
The equation~\eqref{Schrodingerlikeequation1} takes the form
\begin{equation}
\tilde{\psi}''(u)=\lambda_{0}(u)\tilde{\psi}'(u)+s_{0}(u)\tilde{\psi}(u),\label{Schrodingerlikeequation2}
\end{equation}
where
\begin{eqnarray}
	\lambda_{0}(u)&=&-\frac{2 u \left(u^4+2 i \left(u^2+1\right) \frac{m}{k} +2 u^2-3\right)}{\left(u^2-1\right)^2 \left(u^2+1\right)},\label{lambda0}\\
	s_{0}(u)&=&\frac{1}{4 \left(u^2+1\right) \left(u^6-2 u^4+2 u^2-1\right)^2}\nonumber\\
	&&\times \Bigg[3 \left(6 u^4-7 u^2+6\right) \left(u^2+1\right)^3\nonumber\\
	&&+8 i \left(u^2-1\right) \left(u^4-u^2+1\right)^2 \frac{m}{k} \nonumber\\
	&&-4 \left(u^4-u^2+1\right)^2 \left(u^2+1\right)\frac{m ^2}{k ^2}\Bigg].\label{s0}
\end{eqnarray}
With $\lambda_{0}$ and $s_{0}$, we can construct expressions for $c_{n}^{i}$ and $d_{n}^{i}$. Then by the ``quantization condition"~\eqref{QNMscondition3}, we can obtain the QNF of graviscalar QNMs of the thick brane. The results are shown in Tab.~\ref{tab1} and Fig.~\ref{figqnmsplot}. From Tab.~\ref{tab1}, we can see that the first three QNFs for the AIM are in good agreement with the results of the WKB method, but the fourth is not quite the same. This is because the WKB method is not suitable for solving high overtone modes.

The above results are obtained by choosing $a=0$, but since $\omega^{2}=a^2+m^2$, we can easily know the influence of the separation variable parameter a on the QNF: the real part of $\omega$ increases with $a$, while the imaginary part of $\omega$ decreases with $a$. This means that both the oscillation frequency and lifetime of the graviscalar QNMs on the brane increase with $a$. In addition, the value of $a$ will also significantly affect the tail behavior of the graviscalar QNMs, which will be seen in the next section.

\begin{figure}
	\subfigure[~Gaussian pulse]{\label{figlogevenqnms}
		\includegraphics[width=0.23\textwidth]{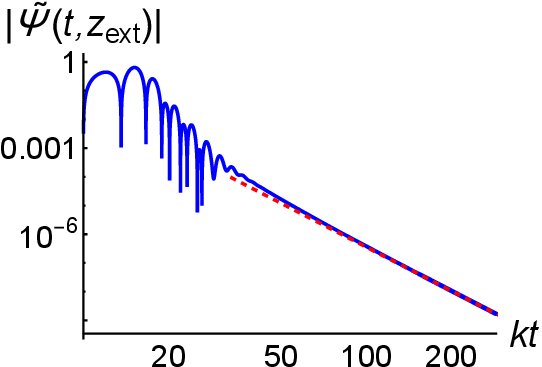}}
	\subfigure[~odd wave packet]{\label{figlogoddqnms}
		\includegraphics[width=0.23\textwidth]{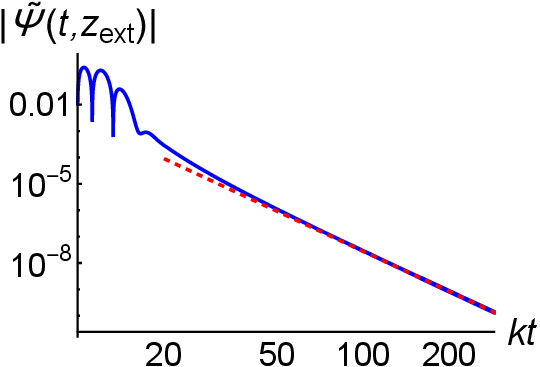}}
	\caption{(a): Time evolution of the Gaussian pulse at $kz=10$. (b): Time evolution of the odd wave packet at $kz=10$.}\label{extqnmmodefig}
\end{figure}

$Time~evolution$ 
Both AIM and WKB methods study the graviscalar QNFs of the thick brane in the frequency domain. These frequencies tell us how perturbations decay in the late time. However, this lacks an understanding of the relative amplitudes of these fluctuations. Therefore, we should study the time evolution of the perturbed scalar field. Here, we use the null coordinates $u=t-z$ and $v=t+z$ to perform the time evolution. The evolution equation~\eqref{evolutionequation} in the null coordinates is
\begin{eqnarray}
	\left(4\frac{\partial^{2}}{\partial u\partial v}+U+a^2\right)\tilde{\Psi}=0. \label{uvevolutionequation}
\end{eqnarray}
Unlike the case of a black hole, the effective potential of the scalar perturbation of the thick brane is symmetric. Thus, the KK modes is either even or odd. Therefore, initial wave packets with different parity will excite different QNMs. It's important to note that the cases where $a=0$ and $a\neq0$ are different. The former corresponds to the evolution of a massless field in the background spacetime, while the latter is the evolution of a massive field. We first consider the case where $a=0$. The initial date is assumed to be a Gaussian pulse first
\begin{eqnarray}
	\tilde{\Psi}(0,v)=e^{\frac{-(kv-kv_{c})^{2}}{2k^2\sigma^{2}}}, ~~~\tilde{\Psi}(u,0)=e^{\frac{-k^2v_{c}^{2}}{2k^2\sigma^{2}}}.\label{gausspulseinitialwavepackage}
\end{eqnarray}
This Gaussian pulse is located at $kv_{c}=5$ and has a width of $k\sigma=1$. We extract the data at $kz_{\text{ext}}=10$. The result of the 
time evolution of the Gaussian pulse is shown in Fig.~\ref{figlogevenqnms}. It can be seen that in logarithmic coordinates, the waveform is divided into three stages: the initial burst stage, the exponential damping stage, and the power-law tail stage. Since the QNM with the smallest imaginary part dominates the evolution, we can extract the frequency of the QNM with the smallest imaginary part by fitting the data. The fitted QNF for the case of Fig.~\ref{figlogevenqnms} is $\frac{m}{k}=1.999394-0.503601i$, which is in good agreement with the result of the AIM and the WKB method.  In addition to the first QNF, we can also extract the frequency of the second QNM due to the symmetry of the effective potential. When the initial data is given as a Gaussian wave packet, both odd and even QNMs are excited. The first QNM is even, and the odd QNM is covered. If the initial data is given as a static odd wave packet, only odd QNMs will be excited. In this way, we can extract the frequency of the first odd QNM. The odd initial data is given by
\begin{eqnarray}
	\tilde{\Psi}(0,v)=\sin\left(\frac{kv}{2}\right)e^{\frac{-k^2v^{2}}{4}}, \\
	\tilde{\Psi}(u,0)=\sin\left(\frac{ku}{2}\right)e^{\frac{-k^2u^{2}}{4}}.\label{oddinitialwavepackage}
\end{eqnarray}
We extract the data at $kz_{\text{ext}}=10$. The result is shown in Fig.~\ref{figlogoddqnms} and the fitted QNF is $\frac{m}{k}=1.576312-1.746200i$. This is also consistent with the result of the AIM and the WKB method. For the high overtone modes, we look forward to developing more methods in the future to compare them with the results of the AIM. 

\begin{figure}
	\subfigure[~Gaussian pulse, $a=\frac{1}{5}$]{\label{figlogevenqnmsm1f5}
		\includegraphics[width=0.23\textwidth]{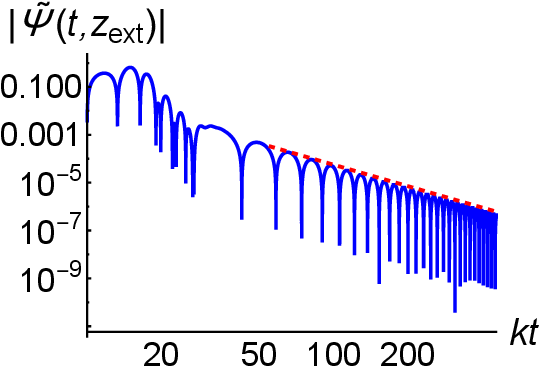}}
	\subfigure[~Gaussian packet, $a=\frac{1}{3}$]{\label{figlogevenqnmsm1f3}
		\includegraphics[width=0.23\textwidth]{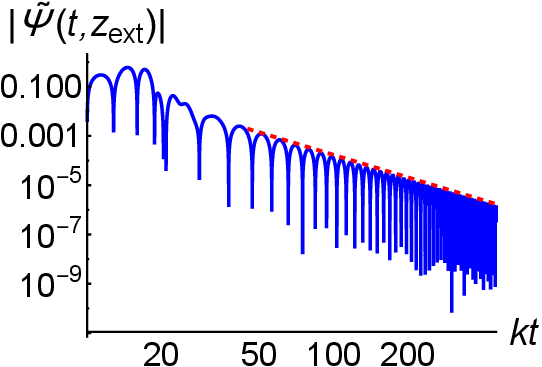}}
	\subfigure[~odd wave packet, $a=\frac{1}{5}$]{\label{figlogoddqnmsm1f5}
		\includegraphics[width=0.23\textwidth]{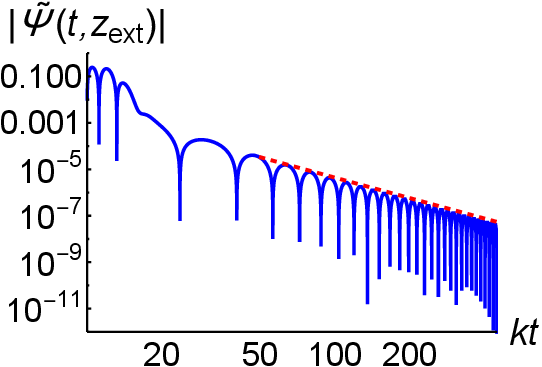}}
	\subfigure[~odd wave packet, $a=\frac{1}{3}$]{\label{figlogoddqnmsm1f3}
		\includegraphics[width=0.23\textwidth]{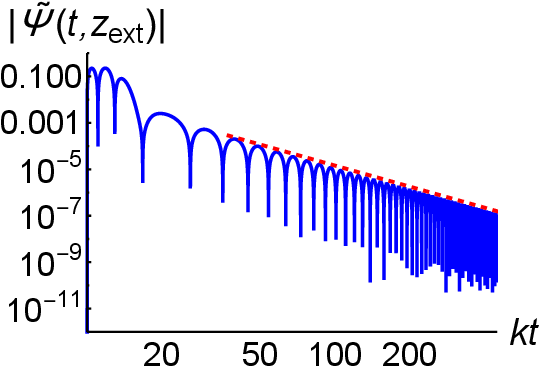}}
	\caption{Upper panel: Time evolution of the Gaussian pulse with different $a$ at $kz_{\text{ext}}10$. Lower panel: Time evolution of the odd wave packet with different $a$ at $kz_{\text{ext}}=10$.}\label{extqnmmodefigmassive}
\end{figure}

Now let's consider the case of $a\neq0$. The selection of the initial wave packet is consistent with the case of $a=0$. The values $a=\frac{1}{5}$ and $\frac{1}{3}$ are selected to study the influence of parameter $a$ on the evolution of the wave packet, and the results are shown in Fig.~\ref{extqnmmodefigmassive}. It can be seen that, when $a\neq0$, there is no power-law tail in the late time, but an oscillating tail. This is consistent with the evolutionary behavior of a massive field in a black hole background. Obviously, the tails for $a=0$ and $a\neq0$ behave completely differently. In a very complete analysis, Ching et al. \cite{Ching:1994bd,Ching:1995tj} examined the late time tail that arises when dealing with this form of the evolution equation~\eqref{evolutionequation} with $a=0$, and the form of effective potential is
\begin{eqnarray}
V(x)\sim \frac{\nu(\nu+1)}{x^2}+ \frac{c_{1}\log x+c_{2}}{x^\alpha}, x\rightarrow\infty.
\end{eqnarray}
When $c_{1}=0$, their conclusions are:\\
(1)~If $\nu$ is an integer, the tail is given by a power-law
\begin{eqnarray}
\tilde{\Psi}\sim t^{-\mu}, ~~~\mu>2\nu+\alpha, 
\end{eqnarray}
where $\alpha$ is an odd integer and $\alpha<2\nu+3$.\\
(2)~For the case of $\nu$ is not an integer, the tail is
\begin{eqnarray}
	\tilde{\Psi}\sim t^{-(2\nu+2)}.
\end{eqnarray}
For the effective potential~\eqref{effectivepotentialz}, we can rewrite it as
\begin{eqnarray}
U(z)=\frac{15k^{2}}{4(1+k^{2}z^{2})}+\frac{3}{4(1+k^{2}z^{2})^{2}}.
\end{eqnarray}
So as $z\rightarrow\infty$, the effective potential asymptotic to
\begin{eqnarray}
	U(z)\sim\frac{15k^{2}}{4k^{2}z^{2}}+\frac{3k^{2}}{4(k^{2}z^{2})^{2}}=\frac{\frac{3}{2}(\frac{3}{2}+1)k^{2}}{k^{2}z^{2}}+\frac{3k^{2}}{4k^{4}z^{4}}.
\end{eqnarray}
Obviously this corresponds to the second case above, so the tail of the wave packet evolution should be $\tilde{\Psi}\sim t^{-5}$. Therefore, the tail of the thick brane QNMs is closely related to the structure of the thick brane, especially the behavior of the extra dimension at infinity. Due to the diversity of thick brane solutions, the types of tail should also be very rich, which needs further study. We verify the above results with the data of numerical evolution. We use a  power-law relation $\tilde{\Psi}=t^{\alpha}$ to fit the late time data in Fig.~\ref{extqnmmodefig} and obtain the value of $\alpha$. The fitted result for the case of Fig.~\ref{figlogevenqnms} is $\alpha=-5.07278$, and for the case of Fig.~\ref{figlogoddqnms} is $\alpha=-5.02804$. Obviously, the late time power-law tails of the evolution of two initial wave packets are consistent and same as the results obtained from the previous analysis (up to numerical error).

\begin{eqnarray}
	U(z)\sim\frac{\frac{3}{2}(\frac{3}{2}+1)k^{2}}{k^{2}z^{2}}.
\end{eqnarray}

On the other hand, the situation is different when $a\neq0$. In the case of black holes, the tail of the massive field is generally divided into two parts: intermediate times and asymptotically late times. The latter came much later than the former. The decay law of massive field at intermediate times is~\cite{Hod:1998ra}  
\begin{eqnarray}
\Psi\propto t^{-\nu-3/2}\sin(\mu t),
\end{eqnarray}
where $\mu$ is the mass of the massive field. The decay law of massive field at asymptotically late times is~\cite{Koyama:2000hj}
\begin{eqnarray}
	\Psi\propto t^{-5/6}\sin(\mu t),
	\end{eqnarray} 
independently on the parameter $\nu$. Due to the similarity of the evolution equation of the graviscalar QNMs of the thick brane to the black hole case, these conclusions should also apply to our case. That is, when $a\neq0$ (note that $a$ essentially acts as $\mu$ here), the decay law at intermediate times of the scalar quasinormal mode should be $\Psi\propto t^{-3}\sin(a t)$ due to $\nu=2/3$ and at asymptotically late times is $\Psi\propto t^{-5/6}\sin(\mu t)$. To prove it, we use function $\tilde{\Psi}=t^{\alpha}\sin(\beta t)$ to fit the late time data in Fig.~\ref{extqnmmodefigmassive} and obtain the values of $\alpha$ and $\beta$, the results are shown in Tab.~\ref{tab2}. We can see that, the tails of different initial data are the same. The decay rate of the tail with time is $t^{-3}$, while the oscillation frequency of the tail is consistent with the parameter $a$. These results are consistent with our previous guesses about the decay law at intermediate times. For the tail at asymptotically late times, which need a long evolutionary time to emerge. Due to the limited time of numerical evolution, we have not been able to observe this behavior in the evolving waveform. However, we believe that this tail also exists in the thick brane scenario.

\begin{table}[htbp]
	\begin{tabular}{|c|c|c|}
		\hline
		$\;\;a\;\;$  &
		$\;\;\text{Gauss pulse}\;\;$  &
		$\;\;\;\;\;\;\;\;\text{Odd wave packet}\;\;\;\;\;\;\;$ \\
		\hline
		~   &~ $\alpha$ ~~  $\beta~~$  &$~~~~\alpha$ ~~ $\beta~~$       \\
		1/5   &-2.99746~~ 0.20169      &~~-2.96658~~ 0.20169       \\
		1/3   &-3.03552~~ 0.333637      &~~-3.03131~~ 0.334265  \\
		\hline
	\end{tabular}
	\caption{The tail parameters $\alpha$ and $\beta$ with different values of $a$ and different initial data calculated by fitting the late time data in Fig.~\ref{extqnmmodefigmassive}.\label{tab2}}
\end{table}

The above results show that a series of discrete QNMs exist in the scalar perturbation of the metric of the thick brane, just like the tensor perturbation of the thick brane. These graviscalar QNMs appear as dissipative scalar particles on the brane, corresponding to the scalar degrees of freedom of gravity. In addition, the frequencies of the first QNM of tensor fluctuation is~\cite{Tan:2022vfe}:  $\frac{m_1}{k}=0.997018-0.526362i$. Comparing the first QNM of scalar fluctuation and tensor fluctuation of thick brane, we can find that the real part of the first graviscalar QNM is larger than the tensor QNM, but the imaginary part is smaller than the tensor QNM. This shows that the graviscalar QNMs have a higher oscillation frequencies (the oscillation frequency of the first graviscalar is twice that of the corresponding tensor mode) and a slightly longer lifetime. Moreover, since the original RS-II thin brane does not have scalar fluctuations, the graviscalar QNMs of the thick brane may be a key feature that distinguishes thin brane from thick brane. If such a QNM is detected, it is conclusive evidence for the existence of extra dimensions and  the brane has a thickness. However, due to the short lifetime and extremely high frequency of these modes (when $k=10^{-3}$ eV and $a/k=0$, the lifetime of the first QNM of the thick brane is $\tau\approx10^{-13}$ s, and the oscillation frequency is about $10^{23}$ Hz), although the lifetime increases with $a$, the oscillation frequency also increases. Therefore, it is almost impossible to detect them with existing gravitational wave detectors. Perhaps high-frequency gravitational wave detectors under construction, such as The CERN Axion Solar Telescope~\cite{Ejlli:2019bqj}, could capture traces of them.
Nevertheless, these graviscalar QNMs may play an important role in the early universe by coupling to the 4-dimensional matter field, in particular the evolution of the early universe and the anisotropy of the cosmic microwave background radiation~\cite{Seahra:2005wk,Seahra:2005iq}. On the other hand, when $a\neq0$, these QNMs have a slowly decaying oscillation tail in the late time. Recently, R. A. Konoplya and A. Zhidenko suggests that these tails of massive patterns from extra dimensions may contribute to the gravitational wave background~\cite{Konoplya:2023fmh}. We expect to find signals of them in existing or future gravitational wave background detectors, such as the pulsar timing array.

\section{Conclusions}
\label{sec:conclusions}
In this paper, we investigated the graviscalar QNMs of the thick brane by the WKB method, the AIM, and the time evolution method. The results show that although the scalar fluctuation of the metric has no bound zero mode, there is still a discrete quasinormal spectrum. These QNMs are closely related to the structure of extra dimension. This is a new understanding of the scalar fluctuation of thick branes, which will help us to better study braneworld models.

We first review the thick brane solution and its scalar fluctuations, then we obtain the evolution equation~\eqref{evolutionequation} and the Schr\"odinger-like equation~\eqref{Schrodingerlikeequation} satisfied by the extra dimensional part of the scalar fluctuations by KK decomposition. Based on the Schr\"odinger-like equation, we use the WKB method and the AIM to solve the graviscalar QNFs of the thick brane, which is shown in Tab.~\ref{tab1}. The results obtained by the two methods are quite consistent. It should be noted that the real part of the frequency of the first QNM of the scalar fluctuation is twice that of the frequency of first QNM of the tensor fluctuation, and the imaginary part is not much different. It is shown that the two lifetimes are close, but the oscillation frequency of the first graviscalar QNM is twice that of the tensor QNM. Using the numerical method, we also investigated the evolution of the initial wave packet on a thick brane. We found that wave packets with different parity will excite different QNMs, because the effective potential is symmetric along the extra dimension. This is different with the cases of black holes. By fitting the evolution data, we obtain the frequencies of the first two QNMs, and the results are in agreement with those obtained in the frequency domain. This shows that our results are credible. Finally, we studied the late time tail of graviscalar QNMs in detail. For the case of $a=0$, similar to a massless field around a black hole, the late time tail of the QNMs of the thick brane is also a power-law. We also find that the tail of different initial wave packet excitation are the same, and the numerical fitting results are consistent with those calculated by Green's function. For the case of $a\neq0$, the situation is similar to a massive field around a black hole. Remarkably, these QNMs, although short-lived, have the potential to play an important role in the early universe. Moreover, for QNMs where $a\neq0$, their slowly decaying oscillating tails have the potential to form the   gravitational wave background. Therefore, there may be imprints of these modes in the gravitational wave background, and it may be worth looking for them in background gravitational wave detectors.

Our work could be improved in several ways. For example, the graviscalar QNMs of other thick branes with more abundant structures, such as $f(R)$-branes, can be studied. The effect of these graviscalar QNMs on the brane is also worthy of further study. 

\vspace{5mm}
\begin{acknowledgments}

This work was supported by the National Natural Science Foundation of China (Grants No. 12035005, No. 12405055, and No. 12347111), the China Postdoctoral Science Foundation (Grant No. 2023M741148), the Postdoctoral Fellowship Program of CPSF (Grant No. GZC20240458), and the National Key Research and Development Program of China (Grant No. 2020YFC2201400).

\end{acknowledgments}

\end{document}